\documentclass[prl,amsmath,twocolumn, showpacs, superscriptaddress,10pt]{revtex4-1}

\usepackage{amsmath}
\usepackage{hyperref}
\usepackage{graphicx}

\expandafter\let\csname equation*\endcsname=\relax
\expandafter\let\csname endequation*\endcsname=\relax

\usepackage{enumerate}
\usepackage{mathbbol}
\usepackage{amsfonts}
\usepackage{color}
\definecolor{Blue}{rgb}{0.00, 0.00, 1.00}
\definecolor{Red}{rgb}{1.00, 0.00, 0.00}

\def\XXint#1#2#3{{\setbox0=\hbox{$#1{#2#3}{\int}$}
     \vcenter{\hbox{$#2#3$}}\kern-.5\wd0}}

\begin{document}

%\title{Random walks and L\'evy flights with stochastic resetting} 
\title{Exact statistics of record increments of random walks and L\'evy flights}

\author{\surname{Claude} Godr\`eche}
\affiliation{Institut de Physique Th\'eorique, Universit\'e Paris-Saclay, CEA and CNRS, 91191 Gif-sur-Yvette, France}
\author{Satya N. \surname{Majumdar}}
\affiliation{LPTMS, CNRS, Univ. Paris-Sud, Universit\'e Paris-Saclay, 91405 Orsay, France}
\author{Gr\'egory \surname{Schehr}}
\affiliation{LPTMS, CNRS, Univ. Paris-Sud, Universit\'e Paris-Saclay, 91405 Orsay, France}

\begin{abstract}

We study the statistics of increments in record values in a time series 
$\{x_0=0,x_1, x_2, \ldots, x_n\}$ generated by the positions of a random 
walk (discrete time, continuous space) of duration $n$ steps. 
%
%Denoting by $R_1 < R_2 < \ldots < 
%R_{M}$ the ordered sequence of the records of the RW, $M$ being the 
%number of records, we focus on the statistics of the record increments 
%$r_k = R_{k+1} - R_{k}$, with $k \geq 1$. 
%
For arbitrary jump length distribution, including L\'evy flights, we show that
the distribution of the record increment becomes {\it stationary}, i.e., independent
of $n$ for large $n$, and compute it explicitly for a wide class of jump distributions.  
In addition, we compute exactly the probability $Q(n)$ that the record increments
decrease monotonically up to step $n$. Remarkably, $Q(n)$ is universal (i..e., independent
of the jump distribution) for each $n$, decaying as $Q(n) \sim {\cal A}/\sqrt{n}$ for large $n$, with a 
universal amplitude ${\cal A} = e/\sqrt{\pi} = 1.53362\ldots$.   
%In particular, we compute 
%explicitly the probability $Q(n)$ that the record increments $r_k$'s 
%are monotonically decreasing up to step $n$ and show that it is completely 
%universal, {\it i.e.}, independent of the jump distribution, for any finite 
%$n$. For large $n$, we show that $Q(n) \sim {\cal A}/\sqrt{n}$, with a 
%universal amplitude ${\cal A} = e/\sqrt{\pi} = 1.53362\ldots$.
%
%We also compute the marginal distribution of the increments and show that it has a very rich structure, depending on the jump distribution. Our analytical results are verified by numerical simulations.  
\end{abstract}

\maketitle

The study of the statistics of records in a time series is fundamental and 
important in a wide variety of systems, including climate studies \cite{hoyt,benestad,basset, RP2006,WK2010,AB2010}, finance 
and economics \cite{records_finance,WBK2011}, hydrology \cite{records_hydrology}, sports \cite{Gembris,sports} and others \cite{glick,gregor_review}. Consider any generic time 
series of $n$ entries $\{x_1, x_2, \ldots, x_n\}$ where $x_i$ may 
represent the daily temperature in a given place, the price of a stock or 
the yearly average water level in a river. A record happens at step $k$ if 
the $k$-th entry exceeds all previous entries, i.e., $x_k > \max\{x_1, 
\ldots, x_{k-1}\}$. Typical questions of interest concern the number of 
records in a given sequence of size $n$, the ages of the records (how long 
a record survives before it gets broken by the next one ?), etc. Another 
natural question is how the actual record value evolves with time $n$. For 
instance, in the context of global warming \cite{basset,RP2006,WK2010}, it is vital to know by how 
much a record temperature gets exceeded by the next record, in other 
words, what are the statistics of the {\it increments} in the record 
values? The increment is the analogue of a `derivative' in 
the sequence of records and it provides important information on the trend of the record sequence, e.g. in the
context of global warming.

\begin{figure}
\includegraphics[width = \linewidth]{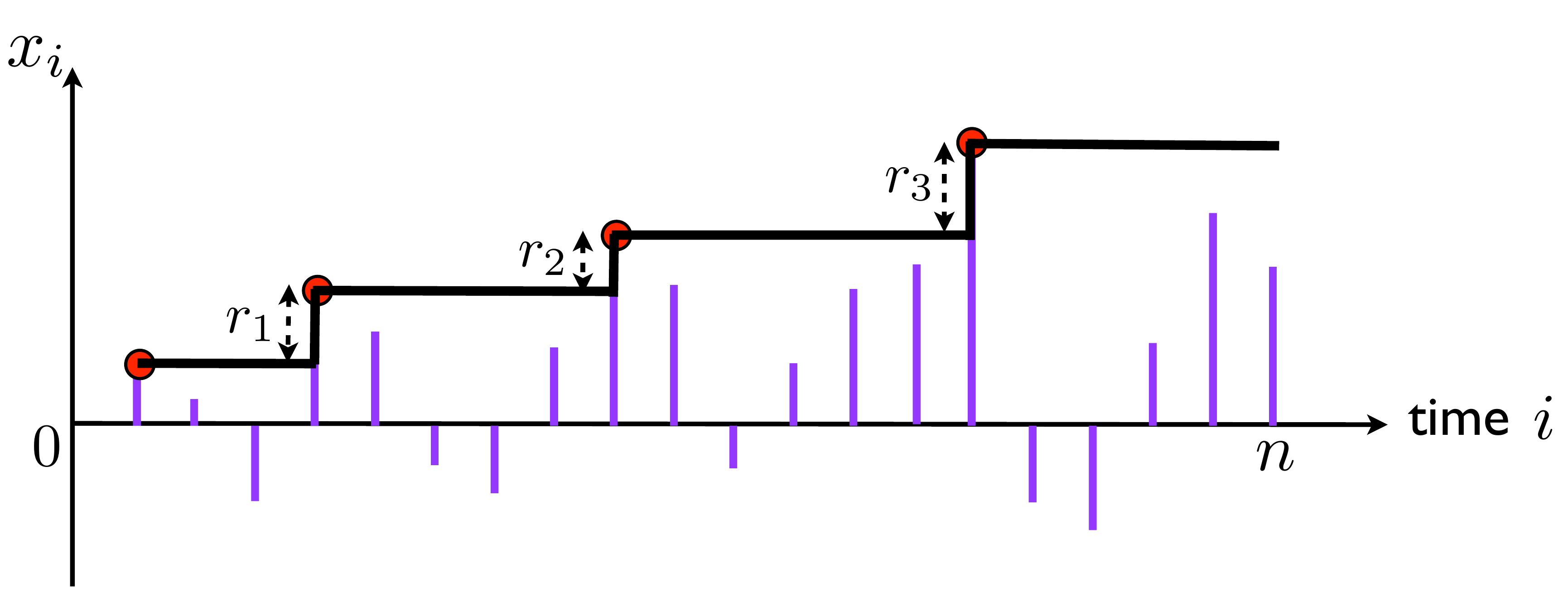}
\caption{For any time series $\{x_1\ldots, x_n \}$, the solid black line representing the current record value as a function of time
exhibits a generic staircase evolution. The increments in record values $r_k$'s are shown by the jumps in the staircase.}\label{fig_staircase}
\end{figure}

Remarkably, the study of records have found a renewed interest and 
applications in diverse complex systems such as the evolution of the 
thermo-remanent magnetization in spin-glasses \cite{jensen,sibani}, evolution of the vortex 
density with increasing magnetic field in type-II disordered 
superconductors \cite{jensen,oliveira}, avalanches of elastic lines in a disordered medium \cite{fisher,sibani_littlewood,ABBM,LDW09}, the evolution of fitness in biological populations \cite{sibani_fitness,krugjain,franke}, and in models of growing networks \cite{GL2008}, amongst others. The 
common feature in all these systems is a {\em staircase} type temporal evolution of relevant observables. For instance, when a domain wall 
in a disordered medium is driven by an increasing
external magnetic field, its center of 
mass remains immobile (pinned by disorder) for a while and then, as the 
field increases further, an extended part of the wall gets depinned, 
giving rise to an avalanche and, consequently, the center of mass jumps 
over a certain distance \cite{fisher,sibani_littlewood,ABBM,LDW09}. The position of the center of 
mass as a function of time (or increasing drive), displays a 
staircase structure as in Fig.~\ref{fig_staircase}. Such a staircase evolution in these 
various systems can be understood in terms of the dynamics of records
in a time series \cite{sibani,sibani_littlewood,oliveira}, where the 
record value remains fixed for a while till it gets broken by the next 
record and jumps by a certain {\it increment} (see Fig.~\ref{fig_staircase}).

Thus the record increments play a crucial role in characterizing the 
generic staircase evolution in such diverse systems and it is important 
to study their statistics in a generic time series. This raises several interesting questions. 
For instance, what is the distribution of a record increment and how does 
it evolve with time? Are the increments at different times correlated? Do 
the increments monotonically decrease with time? The statistics of the 
increments also play an important role in large data analysis, e.g., in 
the characterization of the 
practical criteria to
decide whether a new entry in a time series is a record (or not).
In practice, the record values can be measured only up to a certain precision 
$\delta$ set by the resolution of a detecting instrument~\cite{bala,rounding,edery,krug_last}. If the increment 
is smaller than $\delta$ the new entry is not counted as a record. Hence 
increments also affect the experimentally measured number of records \cite{bala,rounding,edery,krug_last}.

\begin{figure}
\includegraphics[width = \linewidth]{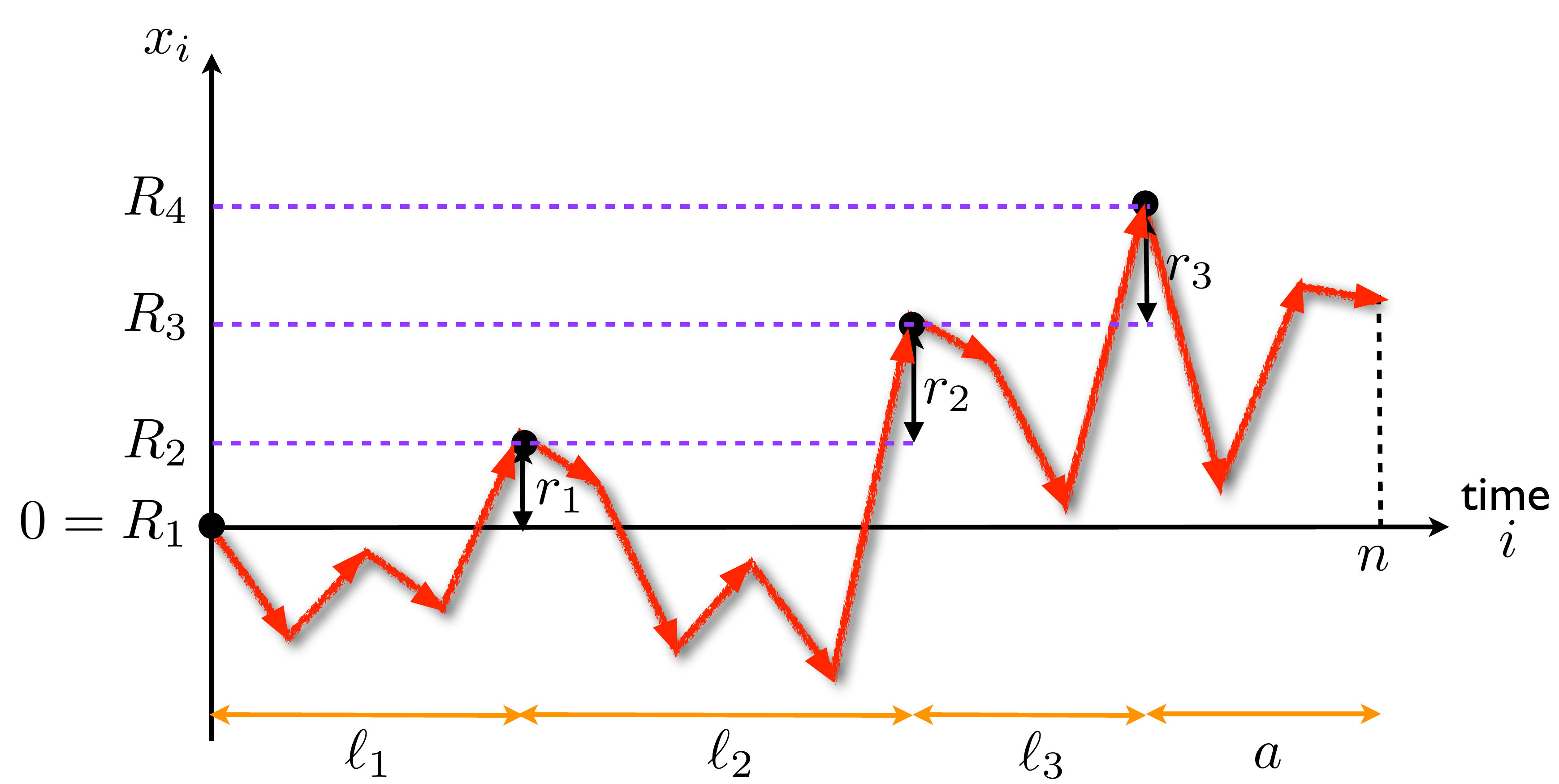}
\caption{Realization of a random walk trajectory of $n=15$ steps with $M = 4$ 
records. The variables $\ell_i$'s denote the ages of the records, i.e., 
the intervals between successive records. The actual record values are 
marked as $R_k$'s and $r_k = R_{k+1}-R_k$ denotes the increment between 
two successive record values.}
\label{fig_intro}

\end{figure}

While the statistics of the total number of records or the time of 
occurrence 
of records have been well studied, 
both for uncorrelated time series 
(where each entry is an independent random variable) \cite{Nevzorov,SM_book2013}, as well as for 
correlated time series such as a random walk (RW) \cite{MZ2008,sanjib2011,SM_book2013,WMS2012,MSW2012,GMS2014}, 
there is hardly any study of the statistics of increments of record values. 
In this Letter, we present exact results for the statistics of increments 
for a RW time series, which
is perhaps the most widely used model of a correlated time 
series in many different 
contexts. For example, in finance, $x_i$ may represent the 
logarithm of the price of a stock \cite{williams} and in queueing  
theory \cite{asmussen}, $x_i$ may mimic the length of a queue at time $i$. 
More precisely, we consider the sequence $\{x_0,x_1,\ldots,x_n\}$ with 
$x_i$ representing the 
position of a random walker on a line
\begin{eqnarray}\label{def_RW}
x_0 = 0 \;, \; x_i = x_{i-1} + \eta_i \;, 
\end{eqnarray} 
where the successive jump lengths  
$\eta_i$ are independent and identically distributed random variables, each drawn from 
a continuous and symmetric probability distribution function 
(PDF) $\phi(\eta)$. Even though the jump lengths are uncorrelated, the 
entries $x_i$'s are strongly correlated. We consider arbitrary 
$\phi(\eta)$, which includes, as special cases,      
the L\'evy flights, where $\phi(\eta) \propto |\eta|^{-1-\mu}$ for large $|\eta|$ (with
$0 < \mu \leq 2$), has a diverging second moment. 
For this model (\ref{def_RW}), the statistics of the total number of records up to step $n$ 
as well as the statistics of the ages of records was shown \cite{MZ2008} to be universal, i.e., independent
of the jump distribution $\phi(\eta)$. Due to strong correlations between the entries, the average
number of records grows as $\sqrt{n}$ for large $n$ \cite{MZ2008},  as opposed to the logarithmic growth for the 
uncorrelated sequence \cite{Nevzorov}. In this Letter, our focus is on the statistics of increments in record values, for
this correlated sequence.

It is useful to summarize our main results. We first compute the full 
joint distribution of the record increments and the total number of 
records for the RW sequence of arbitrary size $n$ and arbitrary jump 
distribution $\phi(\eta)$. One of the important outcomes of this result is 
that the marginal distribution of the record increment becomes {\it 
stationary}, i.e., independent of $n$, for large $n$. However, it does 
depend on the jump distribution $\phi(\eta)$ and we provide explicit 
results for a class of jump distributions. We also compute the probability 
$Q(n)$ that the increments form a monotonically decreasing sequence, an 
important observable studied recently in \cite{MBN13} for an uncorrelated 
sequence. 
Remarkably, we find that $Q(n)$ is totally universal (independent
of $\phi(\eta)$) for {\em each} $n$, despite the fact that 
the increment distribution depends explicitly on $\phi(\eta)$.
Our exact formula reads
\begin{equation}\label{eq:explicit_qn}
Q(n) = e\,\sqrt{\frac{2}{\pi}} \, 
K_{n+1/2}(1) \frac{2^{-n}}{n!} = 
\sum_{j=0}^n {{n+j} \choose n} \frac{2^{-n-j}}{(n-j)!} \;,
\end{equation}
where $K_\nu(x)$ is the modified Bessel function of index $\nu$. 
For instance, $Q(1) = 1$, $Q(2) = 7/8$, $Q(3) = 37/48$, etc.  
For large $n$, we find that $Q(n)$ decays as a power law
\begin{eqnarray}\label{eq:Qn_asympt}
Q(n) \sim \frac{{\cal A}}{\sqrt{n}} \;, \; 
{\cal A} = \frac{e}{\sqrt{\pi}} = 1.53362 \ldots \; ,
\end{eqnarray}
which holds even for L\'evy flights!

We start with a RW sequence in Eq. (\ref{def_RW}). Consider a particular realization with $M$ number of records. 
Let $R_k$'s denote the record values and $r_k = R_{k+1} - R_{k}$ the corresponding increments in this realization (see Fig. \ref{fig_intro}).  
The central object of our computation is the joint probability density 
$P(\vec{r},M,n)$ of the increments $\vec{r}=(r_1, r_2,\ldots, r_{M-1})$ 
and the number of records $M$ for a fixed number of steps $n$ (see Fig. \ref{fig_intro}). To compute 
$P(\vec{r},M,n)$, it turns out 
that we also need to keep track of the record ages $\vec{\ell}\equiv 
\{\ell_1,\ell_2,\ldots, \ell_{M-1}, a\}$ where $\ell_k$ denotes the time 
steps between the $k$-th and $(k+1)$-th record (see Fig.~(\ref{fig_intro})).
Note that the age of the last record is denoted by $a$ 
as it has a slightly different statistics than the preceding ages. This is  because 
the last record, by definition, is still unbroken at the last step $n$ 
while the preceding ones have already been broken. The main idea is that 
one can compute explicitly the `grand' joint PDF $P(\vec{r}, 
\vec{\ell},\,M, \, n)$ of the record increments $\vec{r}$, the record ages 
$\vec{\ell}$, and the number of records $M$ and then integrate out the 
`age' degrees of freedom $\vec{\ell}$ to obtain $P(\vec{r},M,n)$.

To compute $P(\vec{r}, \vec{\ell},\,M, \, n)$, we need 
three quantities as input:

$\bullet$ The first one is the probability $q(\ell)$ that 
a RW, starting at $x_0$, stays below $x_0$ up to $\ell$ time steps:
\begin{eqnarray}\label{def_ql}
q(\ell) = {\rm Prob} (x_i < x_0, \; 1 \leq i \leq \ell) \;,
\end{eqnarray}  
and we define $q(0)=1$.
Due to translational invariance, this 
probability is independent of $x_0$ and we can thus set $x_0=0$. Its 
generating function (GF) is given by the Sparre Andersen 
theorem~\cite{SA53} (for recent reviews see 
~\cite{Redner_book,satya_leuven,Bray_review}):
\begin{eqnarray}\label{SA_th}
\tilde q(z) = \sum_{\ell\geq 0} q(\ell) z^\ell = \frac{1}{\sqrt{1-z}} \Longrightarrow q(\ell) = {2\ell \choose \ell} \frac{1}{2^{2\ell}} \;.
\end{eqnarray}
Note that $q(\ell)$ is completely universal, i.e., independent of $\phi(\eta)$.

$\bullet$ The second quantity we need is the first passage probability $f(\ell)$ of the RW (starting at $x_0=0$)
defined as
\begin{equation}\label{def_fl}
f(\ell) = {\rm Prob} (x_1 < 0, x_2<0, \ldots, x_{\ell - 1}<0, x_{\ell}>0) \;.
\end{equation}  
It follows that $f(\ell) = 
q(\ell-1) - q(\ell)$, so that its GF is also universal, given by
\begin{eqnarray}\label{eq:GF_f}
\tilde f(z) = \sum_{\ell\ge1} f(\ell) z^\ell = 1 - (1-z)\tilde q(z) = 
1 - \sqrt{1-z}\, .
\end{eqnarray}

$\bullet$ Finally, the third quantity we need is $J(\ell,r)$ (for a RW starting at $x_0=0$), defined as
 \begin{equation}\label{def_Jl}
J(\ell,r) =  {\rm Prob} (x_1 < 0, x_2<0, \ldots, x_{\ell - 1}<0, x_{\ell} = r > 0) \;.
\end{equation} 
This denotes the probability that the walker, starting at the origin $x_0=0$, stays below the origin up to $\ell -1$ steps
and then jumps to the positive side, arriving at $r > 0$ at step $\ell$. If one integrates it over the final position $r$, one recovers the first
passage probability at step $\ell$, i.e., 
\begin{eqnarray}\label{eq:identity}
\int_0^\infty J(\ell,r) \,dr = f(\ell) \;.
\end{eqnarray}
The probability $J(\ell,r)$ has also appeared before in the RW literature in different contexts, e.g., in the study of the ordered maxima~\cite{MMS13, MMS14,SM_order,foot_1} and its GF can be computed explicitly in terms of the jump distribution $\phi(\eta)$ (see \cite{supp_mat} for details).

Armed with these three inputs, one can express $P(\vec{r}, \vec{\ell},M,n)$ using the renewal property
of RWs, i.e., the independence of the intervals between successive records (see Fig. \ref{fig_intro}).  
For $M\geq 2$, it reads
\begin{equation}\label{eq:full_joint}
P(\vec{r}, \vec{\ell},M,n) = \prod_{k=1}^{M-1} J(\ell_k,r_k) \, 
q(a) \delta(\ell_1 + \ldots +\,\ell_{M-1}+a,n) \;,
\end{equation}
where $\delta(i,j)$ is the Kronecker delta function, which ensures that 
the total number of steps is fixed to $n$. The factor $q(a)$ corresponds 
to the interval after the last record, {i.e.}, the probability that all 
$x_i$'s after the last record stay below the last record value, which is 
given in Eq. (\ref{SA_th}). For $M=1$, only the starting point is a 
record, and the process stays below $0$ during the entire time interval 
$n$. In this case, there is no record increment, but we set the record 
increment to be $r=0$ by convention and hence
\begin{equation}\label{eq:full_joint_M1}
P(r,a,M=1,n) = q(a) \delta(a,n) \delta(r) \;.
\end{equation}

 The PDF 
$P(\vec{r},M,n)$ is then obtained by summing $P(\vec{r}, \vec{\ell},M,n)$ 
in Eq. (\ref{eq:full_joint}) over $\ell_1, \ldots, \ell_{M-1}$ (each from 
$1$ to $\infty$) and $a$ (from $0$ to $\infty$). Hence the GF of
$P(\vec{r},M,n)$ with respect to $n$ reads, for $M \geq 2$
\begin{eqnarray}\label{eq:gf_jp}
\sum_{n \geq 0} P(\vec{r},M,n) z^n = 
\tilde q(z) \prod_{i=1}^{M-1} \tilde J(z,r_i) \;,
\end{eqnarray}
where $\tilde q(z)$ is given in Eq.~(\ref{SA_th}) and the GF
$\tilde J(z,r)\equiv \sum_{\ell\geq 1} z^\ell J(\ell,r)$. From Eq. (\ref{eq:gf_jp}), it follows that $P(\vec r, M, n)$ is invariant under 
permutation of the labels of record increments, implying that the marginal 
PDF of $r_k$, $P(r_k,n)$, is independent of $k$. It can be computed by 
integrating $P(r,r_2,\ldots,r_{M-1},M,n)$ in Eq. 
(\ref{eq:gf_jp})
over 
$r_2,\ldots,r_{M-1}$ and 
then summing over $M$ (from $1$ to $+\infty$) (see \cite{supp_mat} for 
details). One gets 
\begin{eqnarray}\label{exact_gf_increments}
\sum_{n\geq 0} P(r,n) z^n = \frac{\tilde J(z,r)}{(1-z)} + 
\frac{\delta(r)}{\sqrt{1-z}} \;,
\end{eqnarray}
where we have used $\tilde q(z)=1/\sqrt{1-z}$ [see Eq.~(\ref{SA_th})] and 
$\tilde f(z) = 1- \sqrt{1-z}$ [see Eq.~(\ref{eq:GF_f})]. 
As $z\to 1$, the right hand side of Eq. (\ref{exact_gf_increments}) 
behaves, to leading order, 
as $\tilde J(1,r)/(1-z)$, implying that in the large $n$ limit, 
\begin{eqnarray}
\lim_{n \to \infty} P(r,n) = p(r) = \tilde J(1,r) \;,
\end{eqnarray}
which shows that the PDF of the increments reaches a 
stationary distribution as $n\to \infty$.

%\begin{figure}[t]
%\includegraphics[width = 0.7\linewidth,angle=-90]{plot_pdf_linexp.pdf}
%\caption{
%Marginal distribution of the increments for the jump distribution 
%$\phi(\eta) = |\eta|\,e^{-|\eta|}$ for a RW of $n=10^3$ steps. The points 
%correspond to numerical simulations while the solid line is the exact 
%result obtained in Eq. (\ref{eq:exact_linexp}).}
%\label{fig:linexp}
%\end{figure}
%
\begin{figure}[t]
\includegraphics[width = \linewidth,angle=0]{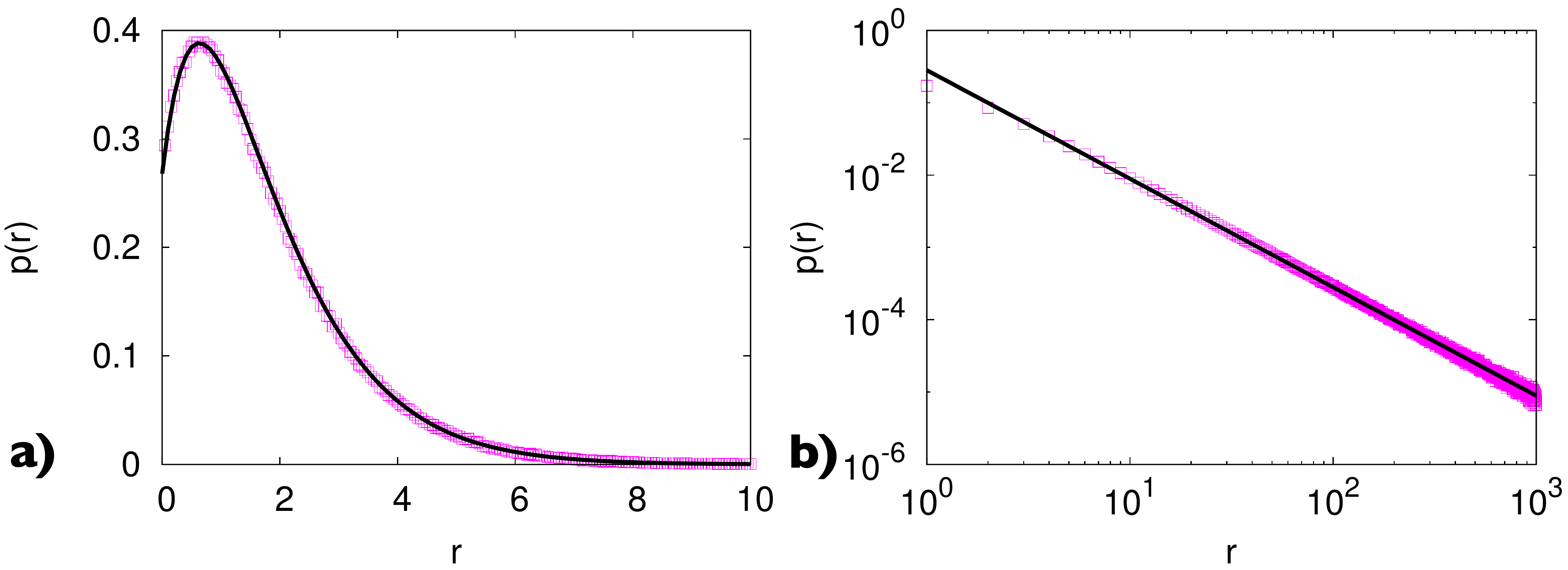}
\caption{
Marginal distribution of the increments for two different jump distributions:
{\bf a)} $\phi(\eta) = (1/2)|\eta|\,e^{-|\eta|}$ and {\bf b)} $\phi(\eta)  = 1/(\pi(1+\eta^2))$ for a RW of $n=10^4$ steps. The square symbols (purple)
correspond to numerical simulations while the solid (black) lines correspond to the exact 
results obtained in Eqs. (\ref{eq:exact_linexp}) in {\bf a)} and (\ref{eq:asympt_heavy}) in {\bf b)}.}
\label{fig:linexp}
\end{figure}
For some jump distributions, $\tilde 
J(1,r)$ can be computed explicitly \cite{supp_mat}. For instance,  
for $\phi(\eta) = (b/2) e^{-b |\eta|}$, one finds $p(r) = b\, e^{-b\,r}$, with  
$r\ge 0$. Another exactly solvable case is $\phi(\eta) = 
(b^2/2) |\eta|\,e^{-b|\eta|}$, for which one finds (with $r\ge 0$)
\begin{eqnarray}\label{eq:exact_linexp}
p(r) = \frac{b^2}{2(1+\sqrt{3})}\,e^{-b\,r}\left(\frac{2}{b} (\sqrt{3}-1) 
+ 4\,r \right) \;.
\end{eqnarray}
In Fig. \ref{fig:linexp} a) we show a comparison between numerical 
simulations and this exact result (\ref{eq:exact_linexp}). The agreement is excellent. For L\'evy flights with $\phi(\eta) \sim A \, |\eta|^{-1-\mu}$ 
with $0<\mu<2$, one can obtain the tail of $p(r)$ exactly for large $r$
\begin{eqnarray}\label{eq:asympt_heavy}
p(r) \sim B_\mu \, r^{-1 - \mu/2} \;, \; r \to \infty \;,
\end{eqnarray}
where $B_\mu$ can be computed explicitly (see Supp. Mat. \cite{supp_mat}). It thus decays more slowly 
than the jump distribution. In Fig.~\ref{fig:linexp}~b) we compare our exact results with
simulation for the Cauchy distribution $\phi(\eta) = 1/(\pi(1+\eta^2))$, corresponding to $\mu =1$. 
Its asymptotic behavior shows a very good agreement with 
our exact result in Eq. (\ref{eq:asympt_heavy}).

We now turn to the computation of $Q(n)$, i.e., the probability that the increments are monotonically
decreasing for the RW 
sequence. To compute $Q(n)$, we first write it as $Q(n) = 
\sum_{M\geq 1}Q(M,n)$ where $Q(M,n)$ is the joint probability that an 
$n$-step RW sequence has exactly $M$ records and that the record 
increments are monotonically decreasing. This probability $Q(M,n)$ is 
obtained by integrating $P(\vec r,M,n)$ over $r_1>r_2>\ldots>r_{M-1}>0$. 
It turns out that these nested integrals can be easily computed from Eq. 
(\ref{eq:gf_jp}) (see \cite{supp_mat}) by the method of induction to yield, 
for any $M \geq 1$
\begin{eqnarray}\label{GF_inter}
\sum_{n\geq 0} z^n Q(M,n) = \tilde q(z) \frac{1}{(M-1)!} \left[\tilde f(z) \right]^{M-1} \;,
\end{eqnarray}
which, quite remarkably, is completely independent of the jump 
distribution $\phi(\eta)$, as $\tilde q(z)$ and $\tilde f(z)$ are themselves 
universal, thanks to the Sparre Andersen theorem. 
\begin{figure}
\includegraphics[width=0.7\linewidth,angle=-90]{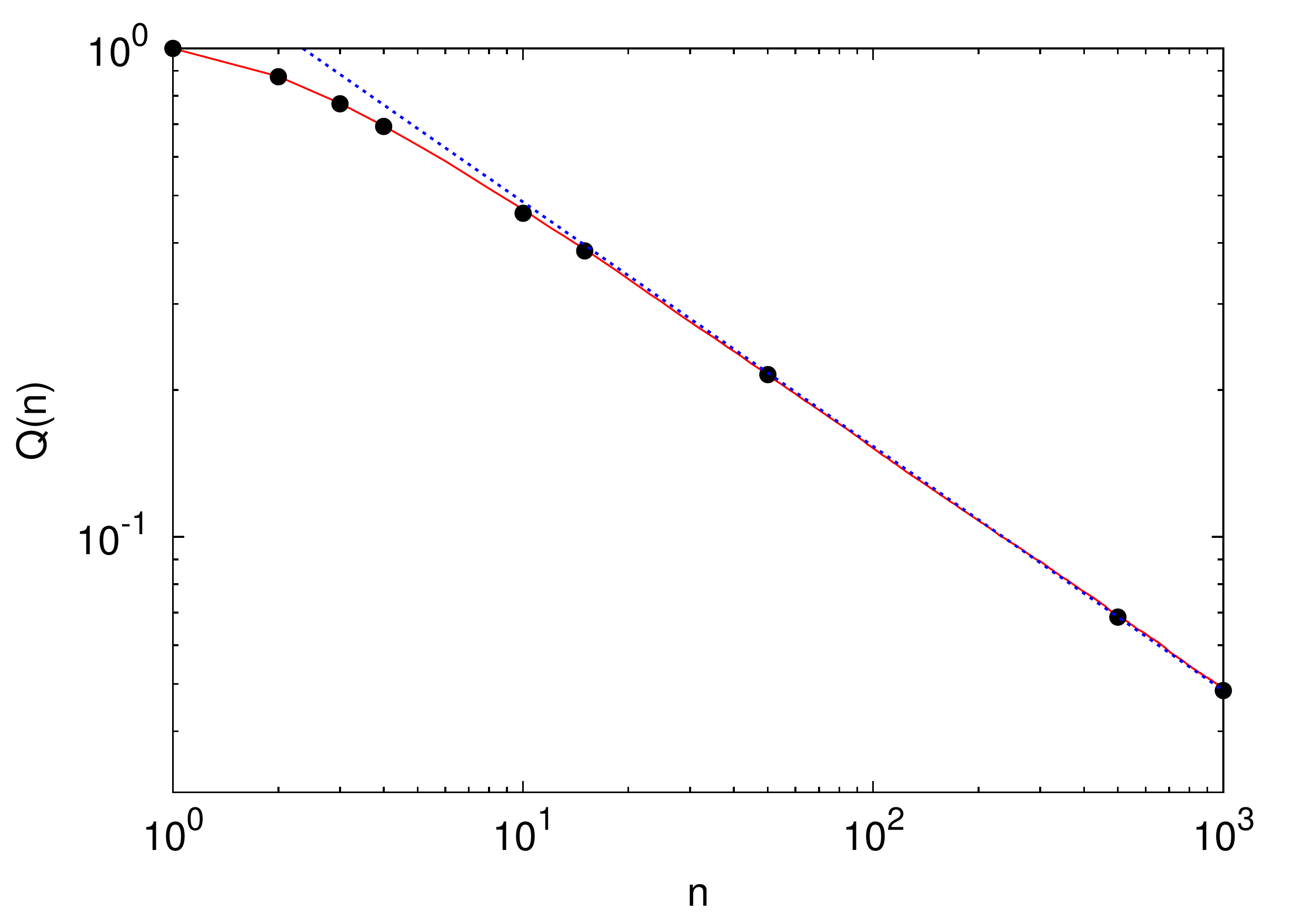}
\caption{$Q(n)$ as a 
function of $n$ for different jump distributions $\phi(\eta)$: uniform, exponential and Cauchy distributions. The three 
curves are indistinguishable, confirming the universality of $Q(n)$, for all $n$. The filled circles 
correspond to the exact finite $n$ value given 
in Eq. (\ref{eq:explicit_qn}). The dotted (blue) line 
indicates the exact asymptotic value 
$Q(n) \sim e/\sqrt{\pi n}$ (\ref{eq:Qn_asympt}).}\label{fig:qn}
\end{figure}

The result in (\ref{GF_inter})
has a nice combinatorial interpretation. From (\ref{eq:gf_jp}), the 
joint PDF $P(r_1, \ldots, r_{M-1},M,n)$ is a symmetric function of $r_k$'s.
Hence, given that the
number of records is exactly $M$, the probability that the sequence
of increments is monotonically decreasing $r_1>\ldots>r_{M-1}$ is just
$1/(M-1)!$ as the $(M-1)!$ possible orderings of these variables are all
equivalent and occur with the same probability. This gives
\begin{eqnarray}\label{relation_Q_Rn}
Q(M,n) = \frac{1}{(M-1)!} F(M,n)\,,
\end{eqnarray}
where $F(M,n)$ is the PDF of the number of records $M$. 
Its GF, $\sum_{n \geq 1} F(M,n) z^n$, was computed in Ref. \cite{MZ2008}
to be exactly ${\tilde q}(z) \left[{\tilde f}(z)\right]^{M-1}$, which
is completely consistent with (\ref{GF_inter}). 
From Eq. (\ref{relation_Q_Rn}), one can  
compute explicitly $Q(M,n)$ (using the result from 
Ref. \cite{MZ2008}):
\begin{eqnarray}\label{eq:qmn_explicit}
Q(M,n) = \frac{2^{-2n + M -1}}{(M-1)!} {2\,n-M+1 \choose n}  \;, 
\end{eqnarray} 
valid for $1 \leq M \leq n+1$ otherwise $Q(M,n) = 0$. Finally, one obtains 
a compact expression of $Q(n)$ by summing up $Q(M,n)$ given in Eq. 
(\ref{eq:qmn_explicit}) over $1 \leq M \leq n+1$, yielding 
Eq. (\ref{eq:explicit_qn}) announced in the introduction \cite{foot_Bessel}. 
To analyze the large $n$ behavior of $Q(n)$, it is convenient to sum
directly over $M$ in Eq. (\ref{GF_inter}) which gives
\begin{eqnarray}\label{eq:Qgf}
\tilde Q(z) = \sum_{n\geq 0} Q(n) z^n = \frac{e^{1-\sqrt{1-z}}}{\sqrt{1-z}} \;,
\end{eqnarray}
where we have used Eqs. (\ref{SA_th}) and (\ref{eq:GF_f}). 
As $z\to 1$, one gets ${\tilde Q}(z) \sim e/\sqrt{1-z}$, implying
that for large $n$, $Q(n)\sim e/\sqrt{\pi n}$, as announced in Eq.
(\ref{eq:Qn_asympt}). In Fig. \ref{fig:qn}, we verify numerically that this result is indeed
universal for three different jump distributions, for any finite $n$, in excellent agreement with our exact result (\ref{eq:explicit_qn}). This
universal behavior of $Q(n)$ for RW sequence is very different from that of the uncorrelated sequence, where, for bounded distribution, $Q(n)$ also decays as a power law for large $n$, $Q(n) \sim n^{-\nu}$, but with an exponent $\nu$ which is non-universal \cite{MBN13}.

To conclude, we have obtained an exact expression for the joint distribution of the record increments
of a RW of $n$ steps (including L\'evy flights), from which the statistics of any observable related to the increments can in principle be computed. Here we calculated the marginal distribution of the increments $p(r)$, which becomes independent of $n$ for large $n$. We also computed the probability $Q(n)$ that the increments are monotonically decreasing up to step $n$. Remarkably, while $p(r)$
is not universal and depends on the jump distribution [see Eqs. (\ref{eq:exact_linexp}) and (\ref{eq:asympt_heavy})], $Q(n)$ is universal for any value of $n$ (\ref{eq:explicit_qn}). Our exact results then provide a benchmark for record increment statistics in a wide variety of problems 
where RW time series is used as a basic model. It will be interesting to study for RW sequence other related 
interesting questions concerning the history of records, such as the fraction of so called superior records, 
that has been studied recently for uncorrelated variables \cite{BNK13}. Finally, it will also be challenging to generalize our results on RW to other strongly correlated time-series. One example is the continuous time random walk (CTRW) \cite{montroll65,zkb83,metzler_review}, for which it is possible to obtain exact results for $Q(n)$ (see Supp. Mat. \cite{supp_mat}).

%In view of recent works on records for RWs, one may naturally wonder how these properties get modified for constrained RWs ({\it i.e.} bridges) \cite{} or %in the presence of a drift \cite{MSW2012}. It would also be interesting to extend these results to other strongly correlated time-series.  

%{\bf Add a few words of conclusion. Mention briefly the case of bridges ?}

%\vspace*{0.5cm}

\vspace*{20cm}

\newpage

\newpage

\begin{widetext}

\begin{large}

\begin{center}

SUPPLEMENTARY MATERIAL

\end{center}
\end{large}

\bigskip

We give the principal details of the calculations described in the manuscript of the Letter.

\section{Computation of the probability $J(\ell,r)$}

For a random walk (RW) sequence as in Eq. (1) of the Letter, with arbitrary jump
distribution $\phi(\eta)$ (continuous and symmetric), we derive an exact
expression for the probability $J(\ell,r)$ defined as
\begin{equation}\label{def_Jl_supp}
J(\ell,r) =  {\rm Prob} (x_1 < 0, x_2<0, \ldots, x_{\ell - 1}<0, x_{\ell} = r > 0) \;.
\end{equation} 
This denotes the probability that the walker, starting at the origin $x_0=0$, stays below the origin up to $\ell -1$ steps
and then jumps to positive side, arriving at $r > 0$ at step $\ell$ (see Fig. \ref{fig_J}) below. If one integrates it over the final position $r$, one recovers the first
passage probability at step $\ell$, i.e., 
\begin{eqnarray}\label{eq:identity}
\int_0^\infty J(\ell,r) \,dr = f(\ell) \;.
\end{eqnarray}
The probability $J(\ell,r)$ has also appeared before in the RW literature in different contexts~\cite{MMS13_supp, MMS14_supp,foot_1_supp}
and its GF can be computed explicitly in terms of the jump distribution $\phi(\eta)$, as demonstrated below.  
\begin{figure}[ht]
\includegraphics[width=0.7 \linewidth]{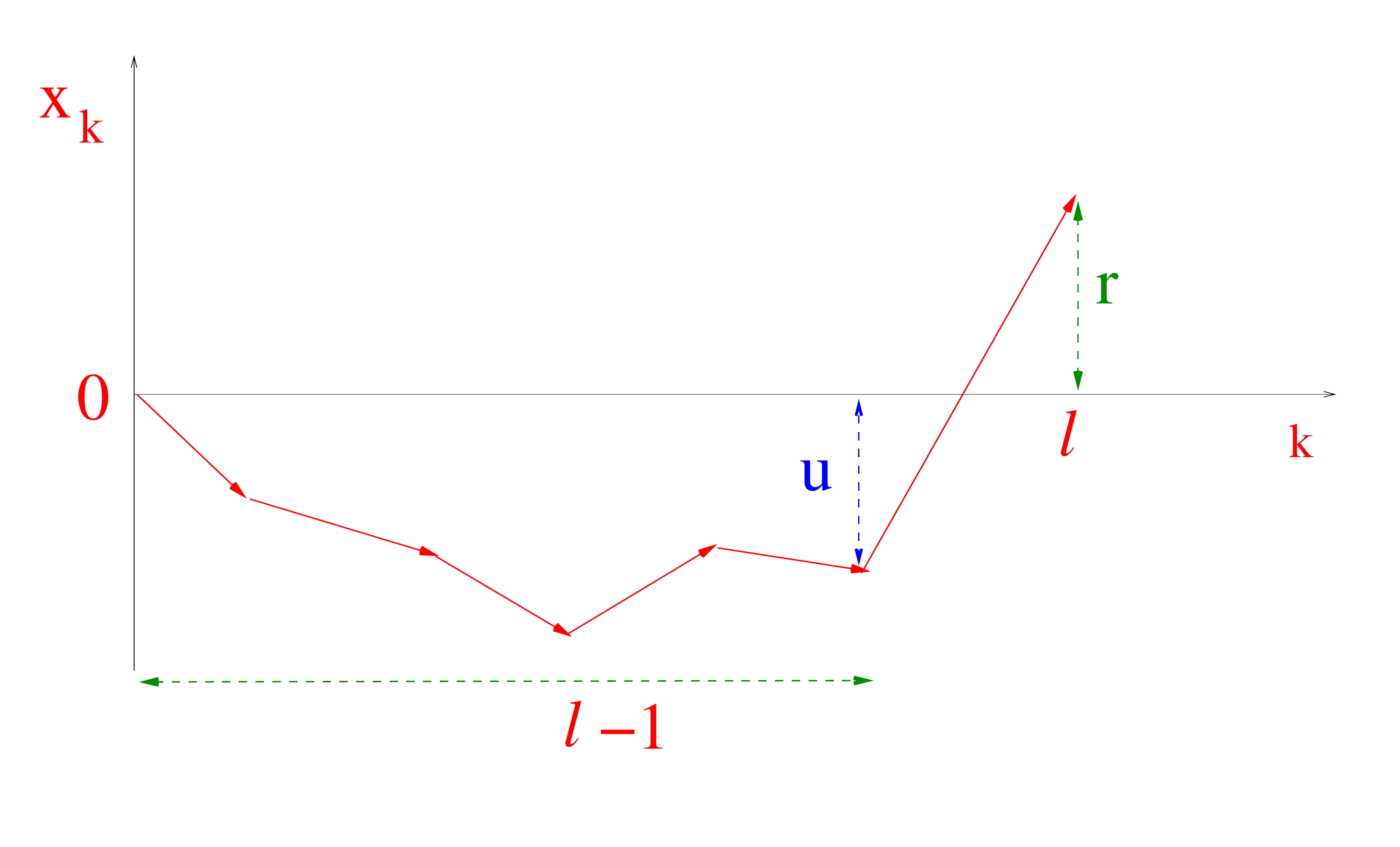}
\caption{A configuration of a RW, starting at the origin $x_0=0$, that stays below the origin up to $\ell - 1$ steps and then jumps to $r>0$ at step $\ell$. We also use the notation $u$  for $ -x_{\ell -1}$, so that the last jump is of length $u+r$.}\label{fig_J}
\end{figure}
To compute $J(\ell,r)$, we first define $G_+(u,\ell-1)$ as the probability density for the walker to arrive at $u>0$ in $\ell-1$ steps, starting from the origin and staying above the origin till $\ell -1$ steps. Note by symmetry $G_+(u,\ell - 1)$ also denotes the probability density that the walker arrives at $-u$ in $\ell -1$ steps, staying negative up to $\ell -1$ steps. Clearly one has
\begin{eqnarray}\label{expr_J}
J(\ell,r) = \int_0^\infty G_+(u,\ell-1) \phi(u+r) \, du \;,
\end{eqnarray}       
where $\phi(u+r)$ denotes the distribution of the last jump (see Fig. \ref{fig_J}) above. Consequently, the GF $\tilde J(z,r) =\sum_{\ell \geq 1}  J(\ell,r) z^\ell$ is given by
\begin{eqnarray}\label{GF_J}
\tilde J(z,r) = \sum_{\ell \geq 0} z^{\ell +1}  \int_0^\infty G_+(u,\ell) \phi(u+r) \, du \;,
\end{eqnarray} 
where we have shifted $\ell$ by 1, for convenience. It turns out that computing the propagator $G_+(u,\ell)$ for arbitrary jump distribution $\phi(\eta)$ is rather nontrivial. Nevertheless there exists, fortunately, an explicit formula \cite{Iva94_supp} for the double Laplace transform of $G_+(u,\ell)$ which reads (for a recent review see \cite{MMS14_supp,satya_leuven_supp})
\begin{eqnarray}\label{eq:ivanov}
\int_0^\infty \sum_{\ell \geq 0} G_+(u,\ell) z^\ell \, e^{-\lambda \, u} \, du = \varphi(\lambda,z) \;.
\end{eqnarray}
The function $\varphi(\lambda,z)$ is given by
\begin{eqnarray}\label{eq:def_phi}
\varphi(\lambda,z) = 
\exp{\left(-\frac{\lambda}{\pi} \int_0^\infty 
\frac{\ln{[1-z\,\hat \phi(k)]}}{k^2+\lambda^2} \, dk \right)} \;,
\end{eqnarray}   
where $\hat \phi(k) = \int_{-\infty}^\infty \phi(\eta)\,e^{i k\eta} \, d\eta$ is the Fourier transform
of the jump distribution. Thus the dependence of $G_+(\ell,u)$ on the jump distribution manifests 
through its Fourier transform $\hat \phi(k)$.

\section{Derivation of the marginal distribution of the increment}

We derive here Eq. (13) of the text. Consider first Eq. (12) of the main 
text with $M\ge 2$. We fix $r_1=r$ in Eq. (12) of the main text and 
integrate over $r_2,\,r_3\,\ldots, 
r_{M-1}$ from $0$ to infinity. Denoting $P(r,M,n)= 
\int_0^{\infty} P(r, r_2, 
r_3,\ldots, r_{M-1})\, dr_2\, dr_3\,\ldots dr_{M-1}$,  
this gives, for $M\ge 2$
\begin{equation}
\sum_{n \geq 0} P(r,M,n) z^n = {\tilde q}(z)\,{\tilde 
J}(z,r)\, \left[\int_0^{\infty} {\tilde J}(z,r_k) dr_k\right]^{M-2}\, .
\label{supp1.1}
\end{equation}
Taking GF of Eq. (9) of the main text with respect to 
$\ell$, one gets, using ${\tilde J}(z,r)= \sum_{\ell \geq 1} z^\ell J(\ell,r)$, 
\begin{equation}
\int_0^{\infty} {\tilde J}(z, r_k) dr_k= \sum_{\ell \geq 1} f(\ell)\, z^\ell= 
{\tilde f}(z)\, .
\label{supp1.2}
\end{equation}
Substituting (\ref{supp1.2}) in (\ref{supp1.1}) gives, for $M\ge 2$,
\begin{equation}
\sum_{n \geq 0} P(r,M,n) z^n=   
{\tilde q}(z)\,{\tilde
J}(z,r)\, \left[{\tilde f}(z)\right]^{M-2}\, .
\label{supp1.3}
\end{equation}
On the other hand, for $M=1$, we use Eq. (11) of the main text. Taking 
GF with respect to $n$ and summing over $a$ gives
\begin{equation}
\sum_{n \geq 0} P(r,M=1,n) z^n= {\tilde q}(z)\, \delta(r)\, .
\label{supp1.4}
\end{equation}
Next, we sum over $M$ ($M=2,3\ldots$) in Eq. (\ref{supp1.3}) and add
also the $M=1$ term in Eq. (\ref{supp1.4}). 
Denoting $P(r,n)= \sum_{M \geq 1} P(r,M,n)$,
we get
\begin{equation}
\sum_{n \geq 0} P(r,n) z^n= \frac{{\tilde q}(z)}{1-{\tilde f}(z)}\, 
{\tilde J}(z,r) + {\tilde q}(z)\, \delta(r)\, .
\label{supp1.5}
\end{equation}
Finally, using the result ${\tilde f}(z)= 1-(1-z)\, {\tilde q}(z)$ with
${\tilde q}(z)=1/\sqrt{1-z}$ (from Eqs. (7) and (6) in the main text),
gives the result
\begin{equation}
\sum_{n \geq 0} P(r,n) z^n= \frac{{\tilde J}(z,r)}{(1-z)}+ 
\frac{\delta(r)}{\sqrt{1-z}}\, ,
\label{supp1.6}
\end{equation}
mentioned in Eq. (13) of the main text. As $z\to 1$, the right hand side of Eq. (\ref{supp1.6}) 
behaves, to leading order, 
as $\tilde J(1,r)/(1-z)$, implying that in the large $n$ limit, 
\begin{eqnarray}
\lim_{n \to \infty} P(r,n) = p(r) = \tilde J(1,r) \;,
\end{eqnarray}
which shows that the PDF of the increments reaches a 
stationary distribution as $n\to \infty$. From Eq. (\ref{GF_J}), one has
\begin{eqnarray}\label{tildeJz1}
\tilde J(1,r) = \sum_{\ell \geq 0} \int_0^\infty G_+(u,\ell) \phi(u+r)\, du \;,
\end{eqnarray}
where the double Laplace transform of $G_+(u,\ell)$ is given in Eqs. (\ref{eq:ivanov}) and (\ref{eq:def_phi}). 
Using these expressions, one can in principle evaluate $\tilde J(1,r)$ for arbitrary jump distribution $\phi(\eta)$. 
In practice, it is very hard though to invert the double Laplace transforms in Eq. (\ref{eq:ivanov}) and then evaluate 
$\tilde J(1,r)$ in Eq. (\ref{tildeJz1}). However, in a different context, namely in the study of the ordered maxima of RWs 
where the same quantity $\tilde J(1,r)$ appears, it was shown that $\tilde J(1,r)$ can be explicitly 
evaluated for a class of jump distributions \cite{MMS13_supp,MMS14_supp}, which leads to the results announced in Eqs. (15-16) of the Letter.  
In particular, for jump distributions with power law tail, $\phi(\eta) \sim A/|\eta|^{\mu +1}$ for large $\eta$ (with $0<\mu<2$), the amplitude $B_\mu$ given in Eq. (16) of the Letter is explicitly given by \cite{MMS14_supp}
\begin{eqnarray}
B_\mu = \frac{1}{2 \,\Gamma\left(1-\frac{\mu}{2}\right)} \sqrt{\frac{\pi\, \mu \, A}{\Gamma(\mu) \sin{\left(\frac{\mu \pi}{2} \right)}}} \;.
\end{eqnarray}

\section{Evaluation of a nested integral}

We want to compute the probability that the record increments are monotonically decreasing for a random walk (RW) sequence of $n$ steps, starting with $x_0=0$
and evolving via Eq. (1) of the main text. To compute this probability, we write
\begin{equation}
\label{supp.2}
Q(n)= \sum_{M \geq 1} Q(M,n)\, ,
\end{equation}
where $Q(M,n)$ is the probability that the sequence has $M$ records and that the record increments are monotonically decreasing, i.e., 
\begin{equation}\label{supp.1}
Q(M,n) = {\rm Prob}\,(r_1> r_2> \ldots > r_{M-1}) \;.
\end{equation}
We can obtain $Q(M,n)$ by
integrating $P(\vec r,M,n)$ in Eq. (12) of the main text, over the domain $r_1>r_2>\ldots >r_{M-1}>0$. Hence, using
Eq. (12) of the main text, one obtains a nested $(M-1)$-fold integral
\begin{equation}
\sum_{n \geq 0} Q(M,n) z^n= {\tilde q}(z)\,
\int_{0}^{\infty} dr_1\, {\tilde J}(z,r_1)\,\int_{0}^{r_1} dr_2 \, {\tilde J}(z,r_2)\,\ldots \, \int_{0}^{r_{M-2}} dr_{M-1}\, 
{\tilde J}(z,r_{M-1})\, .
\label{supp.3}
\end{equation}
To compute the nested integral on the right hand side of (\ref{supp.3}), it is convenient to introduce a new auxiliary integral with a 
variable upper limit, defined as follows
\begin{equation}
I_M(x,z)=  {\tilde q}(z)
\int_{0}^{x} dr_1\, {\tilde J}(z,r_1)\,\int_{0}^{r_1} dr_2 {\tilde J}(z,r_2)\ldots \, \int_{0}^{r_{M-2}} dr_{M-1}\, 
{\tilde J}(z,r_{M-1})\, .
\label{supp.4}
\end{equation}
Then it follows that $\sum_{n \geq 0} Q(M,n) z^n= I_M(\infty,z)$.
To compute $I_M(x,z)$,
we derive Eq. (\ref{supp.4}) with respect to $x$ and obtain a recursion relation
\begin{equation}
\frac{dI_M(x,z)}{dx}= \tilde J(z,x) I_{M-1}(x, z)\, ,
\label{supp.5}
\end{equation}
starting with $I_1(x,z)= {\tilde q}(z)$. The solution of this recursion equation can be easily obtained using the method of induction
and one finds
\begin{equation}
I_M(x,z)= {\tilde q}(z)\,  \frac{\left[\int_0^x {\tilde J}(z,r)\, dr\right]^{M-1}}{(M-1)!}, \quad\quad M\ge 1\, .
\label{supp.6}
\end{equation}
Consequently, we get
\begin{equation}
\sum_{n \geq 0} Q(M,n) z^n= I_M(\infty, z)= {\tilde q}(z)\, \frac{\left[\int_0^\infty {\tilde J}(z,r)\, dr\right]^{M-1}}{(M-1)!}\, .
\label{supp.7}
\end{equation}
Using further the identity in Eq. (9) of the main text, we obtain 
\begin{equation}
\int_0^{\infty}  {\tilde J}(z,r)\, dr= \sum_{\ell \geq 1} f(\ell)\, z^\ell= {\tilde f}(z)\, ,
\label{supp.8}
\end{equation}
where ${\tilde f}(z)=1-\sqrt{1-z}$ is given in Eq. (7) of the main text. Substituting (\ref{supp.8}) in (\ref{supp.7}) provides our main result
\begin{equation}
\sum_{n \geq 0} Q(M,n) z^n={\tilde q}(z)\, \frac{\left[{\tilde f}(z)\right]^{M-1}}{(M-1)!}\, ,
\label{supp.9}
\end{equation}
mentioned in Eq. (17) of the main text.
Summing further over $M$ ($M=1,2\ldots$) and using ${\tilde q}(z)=1/\sqrt{1-z}$ gives the final expression 
\begin{equation}
\sum_{n \geq 0} Q(n)\, z^n= \frac{e^{1-\sqrt{1-z}}}{\sqrt{1-z}}\,,
\label{supp.10}
\end{equation}
given in Eq. (20) of the main text.

\section*{Generalization to continuous time random walks}

In this section, we compute $Q_c(t)$,  denoting the probability that the record increments are monotonically decreasing in a continuous time random walk (CTRW) of duration $t$. The subscript '$c$' refers to CTRW. In CTRW, after every jump, the walker waits at the new position 
for a certain random waiting time $\tau$, before the next jump. The waiting time $\tau$'s are i.i.d. random
variables drawn from a distribution $\rho(\tau)$. The jumps $\eta$'s in space are, as before, also i.i.d. random variables
drawn from the distribution $\phi(\eta)$. Therefore the number of steps $\ell$ in a given time interval $[0,t]$ is a random variable whose
distribution is given by 
\begin{eqnarray}\label{p_of_n}
p(\ell,t) = \int_{0}^{\infty} d\tau_1  \int_{0}^{\infty} d\tau_2 \ldots \int_{0}^{\infty} d\tau_\ell  \prod_{i=1}^\ell \rho(\tau_i) \; \delta(\tau_1 + \tau_2 + \ldots + \tau_\ell - t) \;.
\end{eqnarray}
Its Laplace transform (with respect to $t$) is simply given by
\begin{eqnarray}\label{laplace_pn}
\tilde p(\ell,s) = \int_0^\infty e^{-s t} \, p(\ell,t) \, dt = \left[\tilde \rho(s) \right]^\ell  \;,
\end{eqnarray}
where $\tilde \rho(s)$ is the Laplace transform of the waiting time distribution $\rho(\tau)$. Therefore, the probability of the first return to the origin at time $t$, starting at the origin, is given by
\begin{eqnarray}\label{f_of_t}
f_c(t) = \sum_{\ell \geq 0} f(\ell) \, p(\ell,t) \;,
\end{eqnarray}
where $f(\ell)$ is the probability of first return to the origin for the discrete time RW discussed in the main text. As explained in the text, $f(\ell)$ is universal, i.e., independent of the jump distribution $\phi(\eta)$ and its GF is given by
\begin{eqnarray}
\tilde f(z) = \sum_{\ell \geq 0} f(\ell) \, z^\ell = 1 - \sqrt{1-z} \;.
\end{eqnarray}  
Consequently, taking Laplace transform of Eq. (\ref{f_of_t}) we get (see for instance Ref. \cite{sanjib2011_supp})
\begin{eqnarray}\label{ftilde_expl}
\tilde f_c(s) = \sum_{\ell \geq 0} f(\ell) \, \left[\tilde \rho(s)\right]^\ell = 1 - \sqrt{1 - \tilde \rho(s)} \;.
\end{eqnarray}
Note also that the probability $q_c(t)$ of no return to the origin up to time $t$ is given by $q_c(t) = \int_t^\infty f_c(t') \, dt'$. Consequently, its Laplace transform reads $\tilde q_c(s) = (1 -\tilde f_c(s))/s$. 

With these two ingredients, we now calculate the probability $F_c(M,t)$ that there are exactly $M$ records in time $t$. This is again given by
\begin{eqnarray}\label{Fc}
F_{c}(M,t) = \sum_{\ell \geq 0} F(M,\ell) \, p(\ell, t)\;,
\end{eqnarray}
where $F(M,\ell)$ is the probability of having $M$ records in $\ell$ steps for a discrete time RW, whose GF is given by 
\begin{eqnarray}\label{GF_F}
\sum_{\ell \geq 0} F(M,\ell) z^\ell  = \tilde q(z) \left[ \tilde f(z) \right]^{M-1} \;.
\end{eqnarray}
As discussed in the main text, $F(M,\ell)$ is also universal, i.e., independent of the jump distribution $\phi(\eta)$. Taking Laplace transform of Eq. (\ref{Fc}), using Eqs. (\ref{laplace_pn}) and (\ref{GF_F}) one obtains
\begin{eqnarray}\label{laplace_Fc}
\tilde F_c(M,s) =  \tilde q_c(s) \left[ \tilde f_c(s) \right]^{M-1} \;.
\end{eqnarray} 
Adapting the arguments leading to Eq. (18) in the main text for discrete time RW, it follows that the probability $Q_c(M,t)$ of having $M$ records in time $t$ with monotonically decreasing increments is given by
\begin{eqnarray}
Q_c(M,t) = \frac{1}{(M-1)!} F_c(M,t) \;,
\end{eqnarray} 
and consequently the Laplace transform of $Q_c(t) = \sum_{M \geq 1} Q_c(M,t)$ reads
\begin{eqnarray}\label{final_eq}
\tilde Q_c(s) = \int_0^\infty Q_c(t) e^{-s t} \, dt = \tilde q_c(s)  \sum_{M\geq 1} \frac{\left[ \tilde f_c(s)\right]^{M-1}}{(M-1)!} = \tilde q_c(s) e^{\tilde f_c(s)}  = \frac{\sqrt{1-\tilde \rho(s)}}{s} e^{1 - \sqrt{1 - \tilde \rho(s)}} \;,
\end{eqnarray}
where we have used $\tilde q_c(s) = (1 -\tilde f_c(s))/s$ and Eq. (\ref{ftilde_expl}). This is the main result of this subsection. We see that $Q_c(t)$, while being independent of the jump distribution $\phi(\eta)$, depends explicitly on the waiting time distribution $\rho(\tau)$. One can easily derive the asymptotic late time decay of $Q_c(t)$ as follows.

At late times, we need to analyze the small $s$ behavior of the Laplace transform $\tilde \rho(s)$.
For $0 < \alpha < 1$ it behaves as
$\tilde \rho(s) = 1 - (\tau_0 s)^\alpha+\dots$, where $\tau_0$ is a microscopic time scale. 
The waiting time distribution $\rho(\tau)$ has a power law tail for large $\tau$, $\rho(\tau) \propto 1/\tau^{1+\alpha}$ (with a divergent mean waiting time). In this case, using (\ref{final_eq}), we obtain
\begin{eqnarray}
\tilde Q_c(s) \sim e \, \tau_0^{\alpha/2} \, s^{\alpha/2-1} \;.
\end{eqnarray}   
This means that, for large $t$, 
\begin{eqnarray}\label{Qt}
Q_c(t) \sim \frac{{\cal A}_\alpha}{t^{\alpha/2}} \;\;\;\; {\rm where} \;\;\;\; {\cal A}_\alpha = \frac{e\, \tau_0^{\alpha/2}}{\Gamma\left(1-\frac{\alpha}{2}\right)} \;\;\;\; {\rm with} \;\;\;\; 0< \alpha < 1 \;.
\end{eqnarray}
For $\alpha > 1$, $\tilde \rho(s) = 1 - s\langle \tau\rangle + \dots$.
The mean waiting time $\langle \tau \rangle$ is now finite.
Hence the asymptotic behavior of $Q_c(t)$ is independent of $\alpha$ and is given by %the value in Eq. (\ref{Qt}) corresponding to $\alpha = 1$, i.e.,
\begin{eqnarray}\label{Qc_alphageq1}
Q_c(t) \sim \frac{e\, \sqrt{\langle\tau\rangle}}{\sqrt{\pi}} \frac{1}{\sqrt{t}}  \;\;\;\; {\rm for} \;\;\;\; \alpha > 1 \;.
\end{eqnarray}
Note that strictly for $\alpha = 1$, one expects logarithmic corrections to this purely algebraic behavior in (\ref{Qc_alphageq1}).

\end{widetext}


\begin{thebibliography}{10}
\bibitem{foot_1_supp}
The function $\tilde J(z,r)$ is denoted by $I_1(z,r)$ in Refs. \cite{MMS13_supp,MMS14_supp}.
\end{thebibliography}
\end{document}